\documentclass{article}%
\usepackage[top=3cm, bottom=3cm, left=3cm, right=3cm]{geometry}
\usepackage{amsmath}
\usepackage{amsfonts}
\usepackage{amssymb}
\usepackage{graphicx}%
\begin{document}

\title{ Decay of light scalar mesons into vector-photon and into pseudoscalar mesons}
\author{Francesco Giacosa$^{\text{(a)}}$ and Giuseppe Pagliara$^{\text{(b)}}$\\$^{\text{(a)}}$\emph{ Institut f\"{u}r Theoretische Physik, JW
Goethe-Universit\"{a}t,}\\\emph{Max von Laue--Str. 1 \ D-60438 Frankfurt, Germany}\\$^{\text{(b)}}$\emph{ Institut f\"{u}r Theoretische Physik,
Ruprecht-Karls-Universit\"{a}t,}\\\emph{Philosophenweg 16, D-69120 Heidelberg, Germany}}
\maketitle

\begin{abstract}
The decays of a light scalar meson into a vector meson and a photon
($S\rightarrow V\gamma)$ are evaluated in the tetraquark and quarkonium
assignments of the scalar states. A link with the radiative decays
$\phi\rightarrow S\gamma$ is established: experimental results for
$S\rightarrow V\gamma$ will allow to understand if the direct, quark-loop
contribution to $\phi\rightarrow S\gamma$ or the kaon-loop contribution is
dominant. Also strong decays $S\rightarrow PP$ -where $P$ denotes a
pseudoscalar meson- are investigated: the tetraquark assignment works better
than the quarkonium one. It is then also discussed why the tetraquark
assignment is favoured with respect to a loosely bound kaonic molecular
interpretation of $a_{0}(980)$ and $f_{0}(980)$ mesons.

\end{abstract}



\bigskip

\section{Introduction}

Theoretical studies of the reactions $S\rightarrow V\gamma,$ where
$S=f_{0}\equiv f_{0}(980),$ $a_{0}\equiv a_{0}(980)$ and $V=\rho,$ $\omega,$
have been performed recently
\cite{insight,black,escribano,ivashyn,tanja,volkov}. Experimental effort is
on-going at Wasa@Cosy to determine the corresponding decay widths \cite{wasa}.

The aim of this work is to calculate $S\rightarrow V\gamma$ in the tetraquark
($\overline{q}^{2}q^{2}$) and quarkonium ($\overline{q}q$) pictures of the
light scalar mesons (see Refs. \cite{amslerrev,klempt,dynrec} and Refs.
therein) and to establish a link between the already performed an the future
experimental work at KLOE \cite{aloisio,kloe} and the on-going analysis at
COSY, where complementary processes involving the $f_{0}$ and $a_{0}$ mesons
are studied.

In this paper the evaluation of $S\rightarrow V\gamma$ is performed in two
distinct ways, which we briefly describe in the following:

In way (1) we use previous results \cite{lupod} obtained by studying the
processes, precisely measured by the KLOE collaboration, $\phi\rightarrow
f_{0}(980)\gamma\rightarrow\pi^{0}\pi^{0}\gamma$ and $\phi\rightarrow
a_{0}(980)\gamma\rightarrow\pi^{0}\eta\gamma$ . The analysis of \cite{lupod}
was based on the assumption that the direct (i.e. quark-loop driven) decay
mechanism of Fig. 1.a (via derivative coupling of scalar-to-pseudoscalar
mesons) dominates, see the next section for details. The results for
$S\rightarrow V\gamma$ turn out to be large compared to other works
\cite{insight,black,escribano,ivashyn,tanja,volkov} and seemingly unlikely
(especially in the $\overline{q}q$ assignment). In particular, the decay
$\rho\rightarrow\sigma\gamma$ turns out to be sizably larger than the
corresponding experimental value \cite{pdg,cmd2}. Also, using (Sakurai's
version of) Vector Meson Dominance (VMD), the $\gamma\gamma$ decays turns out
to be 2 (3) orders of magnitudes larger than the experimental results, in the
tetraquark (quarkonium) scenario. For all these reasons, the scenario in which
the mechanism in Fig. 1.a dominates the radiative $\phi$ decay is disfavored.
However, its final rejection (or, eventually, resurrection) will be possible
as soon as the experimental results for $S\rightarrow V\gamma$ will be known.

In way (2) we start from $\gamma\gamma$ data of $f_{0}$ and $a_{0}$ reported
in Ref. \cite{pdg}. Using VMD, the decays $S\rightarrow V\gamma$ are
evaluated: smaller results
are obtained, in line with other studies and with the experimental value
$\rho\rightarrow\sigma\gamma,$ both in the tetraquark and the quarkonium
scenarios. These results for $S\rightarrow V\gamma$, if confirmed,
unequivocally show that the mechanism in Fig. 1.a is negligible when studying
$\phi$ decays, in turn meaning that the kaon-loop approach of Fig. 1.b dominates.

Besides radiative decays, an important result of our work comes from the study
of decays into two pseudoscalar mesons of the full nonet $\{f_{0}(600),$
$k(800),$ $f_{0}(980),$ $a_{0}(980)\}$, both in the tetraquark and quarkonium
pictures by using the decay amplitudes measured in Refs. \cite{bugg1,bugg2}.
Quite remarkably, the tetraquark assignment works better than the quarkonium
one: in fact, large decay widths for $f_{0}(600)$ and $k(800)$ are found,
while in the quarkonium assignment small -and thus unphysical- values for the
widths are obtained. Moreover, the scalar mixing angle in the $\overline
{q}^{2}q^{2}$ assignment is compatible with zero, in agreement with the
degeneracy of $f_{0}$ and $a_{0}$ being $\frac{1}{2\sqrt{2}}([u,s][\overline
{u},\overline{s}]+[d,s][\overline{d},\overline{s}])$ and $\frac{1}{2\sqrt{2}%
}([u,s][\overline{u},\overline{s}]-[d,s][\overline{d},\overline{s}])$
respectively. (A large value of the mixing angle would inevitably spoil the
mass degeneracy, which is one of the key features in favour of a tetraquark
interpretation of scalars). On the contrary, in the $\overline{q}q$ assignment
the mixing angle turns out to have a sign which is opposite to the one
expected from basic considerations of the axial anomaly.
The paper is organized as it follows: after a brief recall of radiative $\phi$
decays (Section 2), we turn the attention to the results in the tetraquark
(Section 3) and quarkonium scenarios (Section 4) respectively. Then, in
section 5 further discussions and conclusions are presented.

\section{Recall of $\phi\rightarrow S\gamma$\emph{\ }}

\subsection{General considerations}

As a first step we briefly review the mechanisms of the radiative $\phi$
decays: the reaction $\phi\rightarrow S\gamma,$ where $S=f_{0},$ $a_{0}$ is
followed by a subsequent decay of $S$ into two pseudoscalar mesons. In the
case in which $f_{0},$ $a_{0}$ are tetraquark or quarkonium states (i.e.
correspond to a preformed and preexisting, non-dynamically generated states,
see the discussion in Ref. \cite{dynrec}) the chain $\phi\rightarrow
S\gamma\rightarrow PP\gamma$ where $P=\pi,$ $\eta$ can occur following the two
mechanisms described in Fig. 1: Fig. 1.a corresponds to a direct coupling of
the $\phi$ meson to a photon and $S,$ while in Fig. 1.b $\phi$ couples to
$\gamma$ and $S$ via a kaon-loop. If $f_{0},$ $a_{0}$ are loosely bound
$\overline{K}K$ molecular states, then only the diagram in Fig. 1.b contributes.%


In the case of the $f_{0}$ meson, the general interaction Lagrangian which
describes both mechanisms is given by:
\begin{align}
\mathcal{L}_{int,f_{0}}  &  =c_{\phi f_{0}\gamma}\phi_{\mu\nu}f_{0}F^{\mu\nu
}+c_{f_{0}\pi\pi}f_{0}\left(  \partial_{\mu}\overrightarrow{\pi}\right)
^{2}+d_{f_{0}\pi\pi}f_{0}\overrightarrow{\pi}^{2}+\nonumber\\
&  c_{f_{0}KK}f_{0}\left(  \partial_{\mu}K^{+}\right)  \left(  \partial^{\mu
}K^{-}\right)  +d_{f_{0}KK}f_{0}(K^{+}K^{-})+... \label{lintf0}%
\end{align}
where dots refer to the analogous terms with the neutral kaon states. One has:

(a) The direct coupling $c_{\phi f_{0}\gamma}\phi_{\mu\nu}f_{0}F^{\mu\nu}$
(where $\phi_{\mu\nu}=\partial_{\mu}\phi_{\nu}-\partial_{\nu}\phi_{\mu}$ with
$\phi_{\mu}$ the field of the $\phi$ meson, $F_{\mu\nu}=\partial_{\mu}A_{\nu
}-\partial_{\nu}A_{\mu}$ is the electromagnetic strength tensor) induces the
$\phi$ decay, Fig. 1.a. The $f_{0}$ meson decays in $\pi^{0}\pi^{0}$ via the
parameters $c_{f_{0}\pi\pi}$ and $d_{f_{0}\pi\pi}$.

(b) The $\phi$ couples to $f_{0}$ via a kaon-loop, see Fig. 1.b. Then, the
$f_{0}$ subsequently decays. This mechanism, although large-N$_{c}$
suppressed, has been regarded as dominant by many authors \cite{achasov}.

Similarly in the $a_{0}$ case: a coupling $c_{\phi a_{0}\gamma}\phi_{\mu\nu
}a_{0}F^{\mu\nu}$ describes the point-like (quark-loop driven) coupling of the
$\phi$ mesons to $a_{0}$ and $\gamma,$ and analogous coupling constants
$c_{a_{0}\pi\eta},$ $d_{a_{0}\pi\eta},$ $c_{a_{0}KK},$ $d_{a_{0}KK}$ are
introduced. It is important to stress that in the considered
tetraquark/quarkonium interpretation of $a_{0}$ and $f_{0}$ both mechanisms
necessarily take place. It is however still unsettled which one, if any, is dominant.

On the contrary, in the molecular assignment only the parameters $d_{f_{0}KK}$
and $d_{a_{0}KK}$ do not vanish (all other constants can be set to zero, in
particular $c_{\phi f_{0}\gamma}$). The coupling to $\phi$ and also the
coupling to $\pi\pi$ are driven by kaon loops via a subsequent $\overline
{K}K\pi\pi$ and $\overline{K}K\pi\eta$ coupling \cite{tanja}.

\begin{figure}
\begin{center}
\includegraphics[
trim=0.000000in -0.014239in 0.000000in 0.014239in,
height=2.4984in,
width=5.1361in
]%
{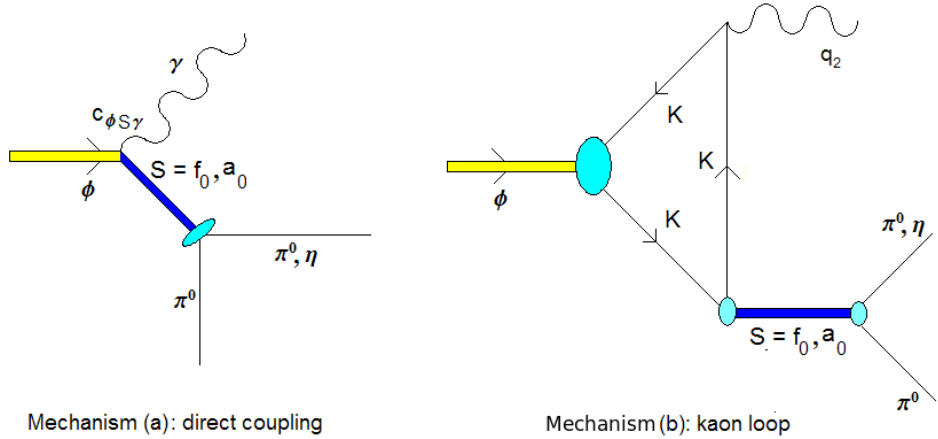}%
\caption{Mechanism (a): direct (quark-loop) contribution. Mechanism (b):
kaon-loop contribution.}%
\label{fig1}%
\end{center}
\end{figure}

\subsection{Fit to line shapes via Fig. 1.a}

In Ref. \cite{lupod}, using the formalism developed in Ref. \cite{lupo}, only
the diagram of Fig. 1.a has been taken into account in the limit $d_{f_{0}%
\pi\pi}=d_{f_{0}KK}=d_{a_{0}\pi\eta}=d_{a_{0}KK}=0$, which corresponds to the
limit in which \emph{only} derivative couplings of scalar to pseudoscalar
mesons are retained. As originally discussed in Ref. \cite{blackradiative},
derivative interactions are potentially interesting because, due to the extra
dependence on the phase space, allow to describe peaked line shapes
$\phi\rightarrow f_{0}(980)\gamma\rightarrow\pi^{0}\pi^{0}\gamma$ and
$\phi\rightarrow a_{0}(980)\gamma\rightarrow\pi^{0}\eta\gamma,$ see details in
\cite{lupod}. By fitting the theoretical curves to the experimental data one
can determine -among others- the coupling constants $c_{\phi f_{0}\gamma}$ and
$c_{\phi a_{0}\gamma},$ which parametrize the point-like coupling. A precise
determination was not possible because the fits depend only marginally on the
couplings to kaons $c_{f_{0}KK}$ and $c_{a_{0}KK}$; for this reasons fits with
different values of the latter have been performed. Nevertheless, a clear and
parameter-independent outcome has been obtained:%
\begin{equation}
c_{\phi f_{0}\gamma}\gtrsim0.25\text{ GeV}^{-1},\text{ }c_{\phi a_{0}\gamma
}\gtrsim0.25\text{ GeV}^{-1}.\label{min}%
\end{equation}
Moreover, when using the strong decay amplitudes obtained in the experimental
works of Refs. \cite{bugg1,bugg2} (see also \cite{bugsummary}):%
\begin{equation}
\left\vert A_{f_{0}\pi\pi}^{\text{exp}}\right\vert =2.88\pm0.22,\text{
}\left\vert A_{f_{0}KK}^{\text{exp}}\right\vert =5.91\pm0.77\text{,
}\left\vert A_{a_{0}\pi\eta}^{\text{exp}}\right\vert =3.33\pm0.15,\text{
}\left\vert A_{a_{0}KK}^{\text{exp}}\right\vert =3.59\pm0.44,\label{amplexp}%
\end{equation}
one finds (tiny errors omitted)
\begin{equation}
c_{\phi f_{0}\gamma}=0.801\text{ GeV}^{-1},\text{ }c_{\phi a_{0}\gamma
}=0.385\text{ GeV}^{-1}.\label{cf}%
\end{equation}
Other choices for the amplitudes are possible by taking the values listed in
the compilation of Refs. \cite{flatte,putative}, where a variety of results
are summarized. They would lead to slightly different values for the
point-like couplings but would not change the qualitative conclusions.

\section{Results in the tetraquark assignment}

\subsection{The Lagrangian}

Be $V_{\mu\nu}=\partial_{\mu}V_{\nu}-\partial_{\nu}V_{\mu}$ with $V_{\mu
}=diag\{\frac{1}{\sqrt{2}}\left(  \rho^{0}+\omega_{N}\right)  ,\frac{1}%
{\sqrt{2}}\left(  -\rho^{0}+\omega_{N}\right)  ,\phi_{S}\}$ the diagonal
matrix with the neutral vector mesons, where $\omega_{N}=\sqrt{\frac{1}{2}%
}(\overline{u}u+\overline{d}d)$ and $\phi_{S}=\overline{s}s.$ The physical
states $\omega$ and $\phi$ are given by%
\begin{equation}
\left(
\begin{array}
[c]{c}%
\omega\\
\phi
\end{array}
\right)  =\left(
\begin{array}
[c]{cc}%
\cos(\varphi_{V}) & \sin(\varphi_{V})\\
-\sin(\varphi_{V}) & \cos(\varphi_{V})
\end{array}
\right)  \left(
\begin{array}
[c]{c}%
\omega_{N}\\
\phi_{S}%
\end{array}
\right)  . \label{phiv}%
\end{equation}
The mixing angle $\varphi_{V}=-3.74^{\circ}$ is close to zero. The quantity
$F_{\mu\nu}=\partial_{\mu}A_{\nu}-\partial_{\nu}A_{\mu}$ is the
electromagnetic strength tensor and $Q=diag\{2/3,-1/3,-1/3\}$ the charge matrix.

The original discussion of tetraquark states as candidates for the light
scalar mesons was presented in Ref. \cite{jaffeorig} and revisited in Ref.
\cite{maiani}. Here we introduce the tetraquark scalar nonet $\mathcal{S}%
^{[4q]}$ using the formalism of Refs. \cite{tq,tqmix}:%
\begin{equation}
\mathcal{S}^{[4q]}=\frac{1}{2}\left(
\begin{array}
[c]{ccc}%
\lbrack\overline{d},\overline{s}][d,s] & -[\overline{d},\overline{s}][u,s] &
[\overline{d},\overline{s}][u,d]\\
-[\overline{u},\overline{s}][d,s] & [\overline{u},\overline{s}][u,s] &
-[\overline{u},\overline{s}][u,d]\\
\lbrack\overline{u},\overline{d}][d,s] & -[\overline{u},\overline{d}][u,s] &
[\overline{u},\overline{d}][u,d]
\end{array}
\right)  =\left(
\begin{array}
[c]{ccc}%
\sqrt{\frac{1}{2}}(f_{B}-a_{0}^{0}) & -a_{0}^{+} & k^{+}\\
-a_{0}^{-} & \sqrt{\frac{1}{2}}(f_{B}+a_{0}^{0}) & -k^{0}\\
k^{-} & -\overline{k}^{0} & \sigma_{B}%
\end{array}
\right)  ,
\end{equation}
where the tetraquark content, in terms of diquark-antidiquark composition, has
been made explicit. Note, the commutator $[.,.]$ reminds that the all diquarks
are in the antisymmetric antitriplet $\overline{3}_{F}$ representation. In
particular, the states $\sigma_{B}[4q]=\frac{1}{2}[u,d][\overline{u}%
,\overline{d}]$ and $f_{B}[4q]=\frac{1}{2\sqrt{2}}([u,s][\overline
{u},\overline{s}]+[d,s][\overline{d},\overline{s}])$ refer to bare (unmixed)
tetraquark scalar-isoscalar states. The physical states $\sigma$ and
$f_{0}\equiv f_{0}(980)$ are then given by%
\begin{equation}
\left(
\begin{array}
[c]{c}%
\sigma\\
f_{0}(980)
\end{array}
\right)  =\left(
\begin{array}
[c]{cc}%
\cos(\varphi_{S}) & \sin(\varphi_{S})\\
-\sin(\varphi_{S}) & \cos(\varphi_{S})
\end{array}
\right)  \left(
\begin{array}
[c]{c}%
\sigma_{B}\\
f_{B}%
\end{array}
\right)  . \label{phis}%
\end{equation}
where $\varphi_{S}$ is the scalar mixing angle.

The pseudoscalar nonet is described, as usual, by the matrix
\begin{equation}
\mathcal{P}=\left(
\begin{array}
[c]{lll}%
\frac{1}{\sqrt{2}}\pi^{0}+\frac{1}{\sqrt{2}}\eta_{N} & \pi^{+} & K^{+}\\
\pi^{-} & -\frac{1}{\sqrt{2}}\pi^{0}+\frac{1}{\sqrt{2}}\eta_{N} & K^{0}\\
K^{-} & \overline{K}^{0} & \eta_{S}%
\end{array}
\right)
\end{equation}
where $\eta_{N}\equiv\sqrt{\frac{1}{2}}(\overline{u}u+\overline{d}d)$ and
$\eta_{S}\equiv\overline{s}s.$ The physical fields arise as $\eta=\eta_{N}%
\cos\varphi_{P}+\eta_{S}\sin\varphi_{P},$ $\eta^{\prime}=-\eta_{N}\sin
\varphi_{P}+\eta_{S}\cos\varphi_{P}.$ The value $\varphi_{P}=-36.0^{\circ}$
\cite{frascati} is used in the following for definiteness; it lies roughly in
the middle of the phenomenological range from $-32^{\circ}$ to $-44^{\circ}$
found in various studies (variation within this range does not imply any
qualitative change in the following).

The interaction Lagrangian involves $\gamma VS$ couplings (parametrized by
$b_{1}$ and $b_{2})$ and the (derivative) scalar-pseudoscalar $SPP$ couplings
(parametrized by $c_{1}$ and $c_{2}$):%
\begin{align}
\mathcal{L}_{\overline{q}^{2}q^{2}}  &  =b_{1}\mathcal{S}_{ij}^{[4q]}Tr\left[
Y^{j}V_{\mu\nu}Y^{i}Q\right]  F^{\mu\nu}-b_{2}\mathcal{S}_{ij}^{[4q]}Tr\left[
Y^{j}Y^{i}V_{\mu\nu}Q\right]  F^{\mu\nu}+\nonumber\\
&  c_{1}\mathcal{S}_{ij}^{[4q]}Tr\left[  Y^{j}\left(  \partial_{\mu
}\mathcal{P}^{t}\right)  Y^{i}\left(  \partial^{\mu}\mathcal{P}\right)
\right]  -c_{2}\mathcal{S}_{ij}^{[4q]}Tr\left[  Y^{j}Y^{i}\left(
\partial_{\mu}\mathcal{P}\right)  \left(  \partial^{\mu}\mathcal{P}\right)
\right]  \label{lagtq}%
\end{align}
where $\left(  Y^{i}\right)  _{jk}=\varepsilon_{ijk}.$ Note, the Lagrangian of
Eq. (\ref{lintf0}) with $d_{f_{0}\pi\pi}=d_{f_{0}KK}=0$ is part of Eq.
(\ref{lagtq}).

The trace structure is such that the terms $b_{1},b_{2}$ ($c_{1},$ $c_{2}$)
describe the $V\gamma$ ($PP$) decays according to Fig. 2.a and 2.b
respectively. In Fig. 2.a a rearrangement and a subsequent decay of the
tetraquark state take place, while in Fig. 2.b one has first a quark-antiquark
annihilation into one (or more) gluon(s) and then the decay. Note, although
the mechanism of Fig. 2.b is suppressed by a factor $N_{c}$ w.r.t. to Fig.
2.a, it has an important role in phenomenology as discussed in Ref. \cite{tq}.%

\begin{figure}
[ptb]
\begin{center}
\includegraphics[
height=1.4486in,
width=6.4636in
]%
{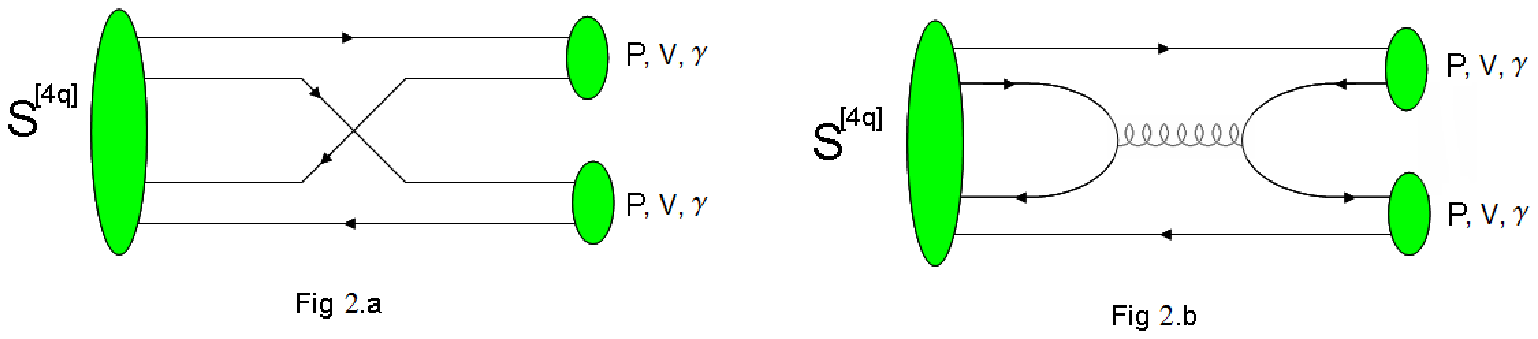}%
\caption{Tetraquark into two pseudoscalars ($PP$), vector and photon
($V\gamma$), or two photons ($\gamma\gamma$) via quark rearrangment (Fig. 2.a)
and annihilation in (at least) one gluon (Fig. 2.b).}%
\label{fig2}%
\end{center}
\end{figure}

\subsection{Strong decays}

The strong decays are parametrized by $c_{1}$ (Fig. 2.a, fall apart decay) and
$c_{2}$ (Fig 2.b). We recall that, in virtue of the derivatives, the
(tree-level) decay widths read:
\begin{equation}
\Gamma_{S\rightarrow P_{1}P_{2}}=\frac{k}{8\pi M_{S}^{2}}\left\vert
A_{SP_{1}P_{2}}\right\vert ^{2},\text{ with amplitude }A_{SP_{1}P_{2}%
}=c_{SP_{1}P_{2}}\frac{(M_{S}^{2}-M_{P_{1}}^{2}-M_{P_{2}}^{2})}{2},
\end{equation}
where $k$ is the three-momentum of an outgoing particle. The coefficients
$c_{SP_{1}P_{2}}$ are obtained by evaluating the traces of Eq. (\ref{lagtq}).
Some relevant ones are reported in Table 1. Most notably, the term
proportional to $c_{2}$ (Fig. 2.b) has two effects: it sizably increases
$f_{B}\rightarrow\overline{K}K$ (thus allowing for a correct description of
the enhanced $f_{B}\rightarrow\overline{K}K$ amplitude in comparison to
$a_{0}\rightarrow\overline{K}K$) and it is responsible for a nonzero
$f_{B}\rightarrow\pi\pi$ amplitude. (Note, in Ref. \cite{thooft} an
instanton-induced term is responsible for a non-zero $f_{B}\rightarrow\pi\pi$
amplitude; here such an instanton term is not needed because the mechanism of
Fig. 2.b is capable of a proper description of phenomenology. Future studies
are required to understand which possibility is realized).

\begin{center}
\textbf{Table 1.} Relevant decay coupling constants

\vspace*{0.5cm}
\begin{tabular}
[c]{|c|c|}\hline\hline
\ $S\rightarrow P_{1}P_{2}$ & $c_{S\rightarrow P_{1}P_{2}}$\\\hline
$f_{B}\rightarrow\overline{K}K$ & $\sqrt{2}\left(  \sqrt{2}c_{1}+\frac
{3}{\sqrt{2}}c_{2}\right)  $\\\hline
$f_{B}\rightarrow\pi\pi$ & $\sqrt{\frac{3}{2}}\cdot\left(  \sqrt{2}%
c_{2}\right)  $\\\hline
$a_{0}\rightarrow\overline{K}K$ & $\sqrt{2}\cdot\left(  \sqrt{2}c_{1}+\frac
{1}{\sqrt{2}}c_{2}\right)  $\\\hline
$a_{0}\rightarrow\pi\eta$ & $-2c_{1}\sin(\varphi_{P})+\sqrt{2}c_{2}%
\cos(\varphi_{P})$\\\hline
$k\rightarrow\pi K$ & $\sqrt{3}\cdot\left(  \sqrt{2}c_{1}+\frac{1}{\sqrt{2}%
}c_{2}\right)  $\\\hline
$\sigma_{B}\rightarrow\pi\pi$ & $\sqrt{\frac{3}{2}}\cdot\left(  2c_{1}%
+2c_{2}\right)  $\\\hline
\end{tabular}

\end{center}

We perform a fit with 3 parameters $c_{1},$ $c_{2}$ and $\varphi_{S}$ to the
four experimental values of Eqs. (\ref{amplexp}). The minimum is found for%
\begin{equation}
c_{1}=5.19\pm1.78\text{ [GeV}^{-1}\text{], }c_{2}=3.84\pm1.80\text{ [GeV}%
^{-1}\text{], }\varphi_{S}=1.2^{\circ}\pm8.0^{\circ};\text{ }\chi
^{2}=1.17.\label{fittq}%
\end{equation}
The small value $\chi^{2}=1.17$ implies that all the amplitudes can be well
reproduced. The scalar mixing angle $\varphi_{S}$ is small and compatible with
zero. This is an important fact: a large scalar mixing angle would spoil the
experimentally well measured degeneracy of $f_{0}$ and $a_{0}.$

As a result of Eq. (\ref{fittq}) one has $\Gamma_{\sigma\rightarrow\pi\pi
}=379\pm52$ MeV, $\Gamma_{k\rightarrow K\pi}=330\pm82$ MeV for masses
$M_{\sigma}=0.6$ GeV and $M_{k}=0.9$ GeV. While the errors refer to the
parameters $c_{1},$ $c_{2}$ and $\varphi_{S}$ only (and are thus
underestimated), it is quite remarkable that large decays of $\sigma$ and $k$
can be obtained from parameters which were fitted in the $f_{0}$ and $a_{0}$
sector \emph{only}. One obtains a qualitative acceptable description of the
full scalar nonet below 1 GeV. This is not the case in the quark-antiquark
assignment, see next sections.

\subsection{Radiative decays}

We first recall that the tree-level decay rates read%
\begin{equation}
\Gamma_{S\rightarrow V\gamma}=\frac{\left(  M_{S}^{2}-M_{V}^{2}\right)  ^{3}%
}{8\pi M_{S}^{3}}g_{SV\gamma}^{2}\text{ , }\Gamma_{V\rightarrow S\gamma}%
=\frac{\left(  M_{V}^{2}-M_{S}^{2}\right)  ^{3}}{24\pi M_{V}^{3}}g_{SV\gamma
}^{2}\text{ .}%
\end{equation}
The couplings $g_{SV\gamma}$ are functions of the two coefficients $b_{1}$ and
$b_{2}$ entering in Eq. (\ref{lagtq}). They are reported in Table 2 where, for
transparency, the limit $\varphi_{V}=0$ and $\varphi_{S}=0$ is considered,
which corresponds to the amplitudes of the bare fields $\omega_{N},$ $\phi
_{S}$ and $\sigma_{B},$ $f_{B}$. Being the mixing angles $\varphi_{V}$ and
$\varphi_{S}$ small, these amplitudes are already close to their real values.
A simple use of the relations of Eqs. (\ref{phiv})-(\ref{phis}) would allow to
obtain the amplitude for the physical states. For completeness we report also
the $\phi$ decay channels, but we stress that, also when assuming that Fig.
1.a is dominant, the tree-level decay rates cannot be used due to the
closeness to threshold and a full study of the line shapes, as for instance
the one in Ref. \cite{lupod}, should be performed.

Independently on the precise value of the parameters, the tetraquark scenario
makes the following predictions: $\Gamma_{a_{0}\rightarrow\omega\gamma}%
\simeq\Gamma_{f_{0}\rightarrow\rho\gamma}$ and $\Gamma_{a_{0}\rightarrow
\rho\gamma}\simeq\Gamma_{f_{0}\rightarrow\omega\gamma.}.$ Also, the mechanism
of Fig. 2.a strongly enhances the (quark-loop driven) decay mode
$\phi\rightarrow f_{0}\gamma$.

\begin{center}
\textbf{Table 2.} $SV\gamma$ decay coupling constants in the $\overline{q}%
^{2}q^{2}$ case

\vspace*{0.5cm}
\begin{tabular}
[c]{|c|c|}\hline\hline
$\phi_{S}$ & $g_{SV\gamma}$\\\hline
$\phi_{B}\rightarrow a_{0}\gamma$ & $\frac{b_{1}}{\sqrt{2}}$\\\hline
$\phi_{S}\rightarrow f_{B}\gamma$ & $\frac{b_{1}+2b_{2}}{3\sqrt{2}}$\\\hline
$\phi_{S}\rightarrow\sigma_{B}\gamma$ & $0$\\\hline
\end{tabular}
\begin{tabular}
[c]{|c|c|}\hline\hline
$\omega_{N}$ & $g_{SV\gamma}$\\\hline
$a_{0}\rightarrow\omega_{N}\gamma$ & $\frac{b_{2}}{2}$\\\hline
$f_{B}\rightarrow\omega_{N}\gamma$ & $\frac{2b_{1}+b_{2}}{6}$\\\hline
$\omega_{N}\rightarrow\sigma_{B}\gamma$ & $\frac{-b_{1}+b_{2}}{3\sqrt{2}}%
$\\\hline
\end{tabular}
\begin{tabular}
[c]{|c|c|}\hline\hline
$\rho$ & $g_{SV\gamma}$\\\hline
$a_{0}\rightarrow\rho\gamma$ & $\frac{2b_{1}+b_{2}}{6}$\\\hline
$f_{B}\rightarrow\rho\gamma$ & $\frac{b_{2}}{2}$\\\hline
$\rho\rightarrow\sigma_{B}\gamma$ & $\frac{b_{1}+b_{2}}{\sqrt{2}}$\\\hline
\end{tabular}

\end{center}

We now calculate the radiative decays in the two ways mentioned in the
Introduction: first, by using the results of Ref. \cite{lupod}, which assume a
dominance of the diagram of Fig. 1.a in radiative $\phi$ decays, and then
starting from $\gamma\gamma$ data of $f_{0}$ and $a_{0}.$ As we shall see,
different results are obtained.

\emph{Way 1:} We fix $\varphi_{S}=1.2^{\circ}$ as determined by the fit to
strong decays. Using Eq. (\ref{cf}) we determine $b_{1}=0.61$ GeV$^{-1}$ and
$b_{2}=1.46$ GeV$^{-1}.$ Note, the ratio $b_{2}/b_{1}\sim2$ is already
problematic because it implies a dominance of the large-N$_{c}$ suppressed
contribution of Fig. 2.b. The results for $S\rightarrow V\gamma$ decays are
summarized in Table 3.

\begin{center}
\textbf{Table 3: }$S\rightarrow V\gamma$ in the $\overline{q}^{2}q^{2}$ case
using Ref. \cite{lupod}.%

\begin{tabular}
[c]{|c|c|}\hline\hline
Mode & Decay [keV]\\\hline\hline
$a_{0}\rightarrow\rho\gamma$ & 535\\\hline\hline
$f_{0}\rightarrow\rho\gamma$ & 1005\\\hline\hline
$\rho\rightarrow\sigma\gamma$ & 876\\\hline
\end{tabular}
\begin{tabular}
[c]{|c|c|}\hline\hline
Mode & Decay [keV]\\\hline\hline
$a_{0}\rightarrow\omega\gamma$ & 1406\\\hline\hline
$f_{0}\rightarrow\omega\gamma$ & 463\\\hline\hline
$\omega\rightarrow\sigma\gamma$ & 20\\\hline
\end{tabular}

Moreover: $\Gamma_{\phi\rightarrow\sigma\gamma}=0.06$ keV (small).
\end{center}

The values in Table 3 are large when compared to the results obtained via
meson-loop contributions, which are typically smaller than 20 keV
\cite{insight,escribano,tanja}. The main point is that large decay widths such
as those in Table 3 are possible only if the direct mechanism of Fig. 1.a dominates.

Both decay modes $\omega\rightarrow\sigma\gamma$ and $\rho\rightarrow
\sigma\gamma$ turn out to be larger than the present experimental knowledge
($\Gamma_{\rho\rightarrow\sigma\gamma}\simeq6$ keV and $\Gamma_{\omega
\rightarrow\sigma\gamma}\leq6$ keV, see Ref. \cite{pdg,cmd2}). While this
represents an argument against this scenario, one should not forget that the
theoretical expressions for these decays strongly depend on the $\sigma$ mass
and on finite-width corrections.

Vector meson dominance can be introduced in the model by applying the shift
$V_{\mu\nu}\rightarrow V_{\mu\nu}+\frac{\sqrt{2}\alpha}{g_{\rho}}QF_{\mu\nu},$
where $g_{\rho}=6.1$ and $\alpha\simeq1/137$ is the fine structure constant.
Although VMD cannot be used for precise calculations of on-shell decays of
photons (see discussion in Ref. \cite{thomas}), it is a valuable
phenomenological tool to estimate the order of magnitude. The results are:
\begin{equation}
\Gamma_{f_{0}\rightarrow\gamma\gamma}\simeq72\text{ keV},\text{ }\Gamma
_{a_{0}\rightarrow\gamma\gamma}\simeq40\text{ keV},\text{ }\Gamma
_{\sigma\rightarrow\gamma\gamma}\simeq26\text{ keV.}%
\end{equation}
These results are sizably larger than the experimental results \cite{pdg}:%
\begin{equation}
\Gamma_{f_{0}\rightarrow\gamma\gamma}^{\text{exp}}=0.29_{-0.9}^{+0.7}\text{
keV},\text{ }\Gamma_{a_{0}\rightarrow\gamma\gamma}^{\text{exp}}=0.3\pm
0.1\text{ keV},\text{ }\Gamma_{\sigma\rightarrow\gamma\gamma}^{\text{exp}%
}=0.5\text{-}4\text{ keV.}\label{ggexp}%
\end{equation}
Thus, the present experimental evidence is against these solutions and thus to
the hypothesis that mechanism of Fig. 1.a plays a dominant role in the $\phi$
decay.

In principle we could also have used the results of the non-structure model of
Ref. \cite{isidori}, in which Fig. 1.a is also regarded as dominant, but
non-derivative interactions of scalar to pseudoscalar mesons are used. The
coupling constants $c_{\phi f_{0}\gamma}$ and $c_{\phi a_{0}\gamma}$
correspond indeed to the lower limit of Eq. (\ref{min}). Smaller results (for
instance $\Gamma_{a_{0}\rightarrow\omega\gamma}\simeq77$ keV) follow, but the
strong amplitudes $A_{SP_{1}P_{2}}$ (fixed by fitting to KLOE line shapes)
determined in Ref. \cite{isidori} turn out to be significantly smaller than
Eq. (\ref{amplexp}). As a consequence, in this scheme the $\sigma$ and $k$
mesons would have a width of $\sim100$ MeV (or even smaller), which is in
clear disagreement with the data.

\bigskip

\emph{Way 2: }We first obtain $b_{1}=0.075$ GeV$^{-1}$ and $b_{2}=0.083$
GeV$^{-1}$ from the $\gamma\gamma$ decays $\Gamma_{f_{0}\rightarrow
\gamma\gamma}^{\text{exp}}=0.29_{-0.9}^{+0.7}$ keV$,$ $\Gamma_{a_{0}%
\rightarrow\gamma\gamma}^{\text{exp}}=0.30\pm0.1$ keV \cite{pdg} (under the
assumptions that they are dominated by the direct, quark-loop in the
tetraquark assignment). Note, in this case $b_{2}/b_{1}\simeq1$ in line with
the ratio $c_{2}/c_{1}$ in the strong sector. As a first consequence a small
$\Gamma_{\sigma\rightarrow\gamma\gamma}=0.14$ keV (which corresponds to the
direct, quark-loop decays and neglects pion loops) is found. Then, via VMD, we
evaluate the $S\rightarrow V\gamma$ decay rates, which are summarized in Table 4.

\begin{center}
\textbf{Table 4: }$S\rightarrow V\gamma$ in the $\overline{q}^{2}q^{2}$ case
using VMD.%

\begin{tabular}
[c]{|c|c|}\hline\hline
Mode & Decay [keV]\\\hline\hline
$a_{0}\rightarrow\rho\gamma$ & 4.0\\\hline\hline
$f_{0}\rightarrow\rho\gamma$ & 3.1\\\hline\hline
$\rho\rightarrow\sigma\gamma$ & 5.1\\\hline
\end{tabular}
\begin{tabular}
[c]{|c|c|}\hline\hline
Mode & Decay [keV]\\\hline\hline
$a_{0}\rightarrow\omega\gamma$ & 4.9\\\hline\hline
$f_{0}\rightarrow\omega\gamma$ & 3.4\\\hline\hline
$\omega\rightarrow\sigma\gamma$ & 0.003\\\hline
\end{tabular}

Moreover: $\Gamma_{\phi\rightarrow\sigma\gamma}=0.004$ keV (very small).

\end{center}

In this case the predicted $\rho\rightarrow\sigma\gamma$ and $\omega
\rightarrow\sigma\gamma$ are in agreement with the experiment \cite{cmd2}. As
a further important consequence, the couplings $c_{\phi f_{0}\gamma}=0.050$
GeV$^{-1}$ and $c_{\phi a_{0}\gamma}=0.054$ GeV$^{-1}$ are determined. They
are factor of 10 \emph{smaller} than the values of Eq. (\ref{cf}). It is
evident that the contribution of Fig. 1.a is reduced of a factor of 100 and is
negligible in this scenario. The only possibility is that the kaon loop of
Fig. 1.b dominates the radiative decay of the $\phi$ mesons.

\section{Results in the quarkonium assignment}

\subsection{The Lagrangian}

A quark-antiquark interpretation of the light scalars is disfavored by the
mass pattern, large $N_{c}$ and various phenomenological arguments (see Refs.
\cite{klempt,noqq} and Refs. therein). However, being not yet fully ruled out,
it is instructive to perform the study within this assignment. The
quark-antiquark nonet is encoded in the matrix
\begin{equation}
\mathcal{S}^{[\overline{q}q]}=\left(
\begin{array}
[c]{lll}%
\overline{u}u & \overline{d}u & \overline{s}u\\
\overline{u}d & \overline{d}d & \overline{s}d\\
\overline{u}d & \overline{d}s & \overline{s}s
\end{array}
\right)  =\left(
\begin{array}
[c]{lll}%
\frac{1}{\sqrt{2}}a_{0}^{0}+\frac{1}{\sqrt{2}}\sigma_{B} & a_{0}^{+} & k^{+}\\
a_{0}^{-} & -\frac{1}{\sqrt{2}}a_{0}^{0}+\frac{1}{\sqrt{2}}\sigma_{B} &
k^{0}\\
k^{-} & \overline{k}^{0} & f_{B}%
\end{array}
\right)  \text{ ,}%
\end{equation}
where the quark content has been made explicit. The scalar-isoscalar states
read $\sigma_{B}=\sqrt{\frac{1}{2}}(\overline{u}u+\overline{d}d)$ and
$f_{B}=\overline{s}s$. The physical states $\sigma$ and $f_{0}\equiv
f_{0}(980)$ arise via mixing:%
\begin{equation}
\left(
\begin{array}
[c]{c}%
\sigma\\
f_{0}(980)
\end{array}
\right)  =\left(
\begin{array}
[c]{cc}%
\cos(\varphi_{S}) & \sin(\varphi_{S})\\
-\sin(\varphi_{S}) & \cos(\varphi_{S})
\end{array}
\right)  \left(
\begin{array}
[c]{c}%
\sigma_{B}\\
f_{B}%
\end{array}
\right)  .
\end{equation}
The Lagrangian for the full-nonet of quark-antiquark, including the
non-derivative flavor symmetry braking term, reads \cite{ecker,longchiral}:%

\begin{equation}
\mathcal{L}_{\overline{q}q}=c_{d}Tr\left[  \mathcal{S}^{[\overline{q}%
q]}\partial_{\mu}\mathcal{P}\partial^{\mu}\mathcal{P}\right]  -c_{m}\frac
{B}{2}Tr[2\mathcal{S}^{[\overline{q}q]}\mathcal{P}M\mathcal{P}+\mathcal{S}%
^{[\overline{q}q]}\mathcal{PP}M+\mathcal{S}^{[\overline{q}q]}M\mathcal{PP}%
]+bTr\left[  \mathcal{S}^{[\overline{q}q]}VQ\right]  F^{\mu\nu}%
\label{lagqqbar}%
\end{equation}
where $M=diag\{m_{u},m_{d},m_{s}\}$ is the diagonal matrix of current quark
masses and the parameter $B$ is related to the pion and kaon masses as
$M_{\pi}^{2}=2Bm_{u}$ and $M_{K}^{2}=B(m_{u}+m_{s}).$

\subsection{Strong decays}

We proceed as in the tetraquark case by performing a fit of the three
constants $c_{d},$ $c_{m}$ and $\varphi_{S}$ to the four experimental
amplitudes of Eq. (\ref{amplexp}). The theoretical expressions for all the
decays can be found in Ref. \cite{longchiral}. Only one acceptable solution is
found:%
\begin{equation}
c_{d}=8.72\pm0.44\text{ GeV}^{-1},\text{ }c_{m}=6.01\pm1.56\text{ GeV}%
^{-1},\text{ }\varphi_{S}=-23.7^{\circ}\pm2.3^{\circ};\text{ }\chi
^{2}=0.02.\label{qfit}%
\end{equation}
Although the $\chi^{2}$ is very small, and thus the amplitudes of Eq.
(\ref{amplexp}) can be correctly reproduced for the above values of the
parameters, one obtains as a consequence that $\Gamma_{\sigma\rightarrow\pi
\pi}=170\pm17$ MeV, $\Gamma_{k\rightarrow K\pi}=218\pm25$ MeV ($M_{\sigma
}=0.6$ GeV and $M_{k}=0.9$ GeV have been used). These decay widths are too
small when compared to experiments. This is a clear drawback of the
quark-antiquark assignment: the parameters obtained from $f_{0}$ and $a_{0}$
resonances do \emph{not} allow for a description of the broad $k$ and $\sigma$
states. Note, as a result of the fit $c_{m}/c_{d}\sim0.7,$ thus sizable. A fit
based only on the chiral symmetric $c_{d}$ term would provide a large
$\chi^{2}.$

A second more subtle but also decisive drawback is the following: a negative
mixing angle $\varphi_{S}$ is a clear outcome of the fit. In the generalized
Nambu Jona-Lasinio model, which provides still one of the main reasons in
favour of a $\overline{q}q$ interpretation of light scalars, the mixing in the
scalar sector is driven by the 't Hooft term which solves the $U_{A}(1)$
anomaly problem in the pseudoscalar sector. In fact, the 't Hooft term induces
the mixing of isoscalar states both in the pseudoscalar and the scalar sectors
and it turns that the mixing angles $\varphi_{P}$ and $\varphi_{S}$ should
have the opposite sign \cite{njl}. Being $\varphi_{P}$ negative, a positive
$\varphi_{S}$ is expected for a quark-antiquark nonet (see also the discussion
in Ref. \cite{longchiral} and Refs. therein). Note, the same conclusion has
been obtained in Ref. \cite{juergen} where a generalized $\sigma$ model with
an anomaly term is studied. Thus, the fact that our fit provides a negative
$\varphi_{S}$ represents a further argument against a quarkonium
interpretation of light scalar mesons. Indeed, a positive $\varphi_{S}$ is the
outcome of Refs. \cite{longchiral}, where the scalar quarkonium nonet is
placed above 1 GeV.

\subsection{Radiative decays}

The theoretical amplitudes for $S\rightarrow V\gamma$ decays in the
$\overline{q}q$ case are reported in Table 5, where $\varphi_{S}$ is kept free
but $\varphi_{V}=0.$ Independently from the way (1) or (2) described later on,
the following ratios are obtained: $\Gamma_{a_{0}\rightarrow\omega\gamma
}/\Gamma_{a_{0}\rightarrow\rho\gamma}\simeq\Gamma_{f_{0}\rightarrow\rho\gamma
}/\Gamma_{f_{0}\rightarrow\omega\gamma}\simeq\Gamma_{\rho\rightarrow
\sigma\gamma}/\Gamma_{\omega\rightarrow\sigma\gamma}\simeq9.$

\begin{center}
\textbf{Table 5:} $SV\gamma$ decay coupling constants in the $\overline{q}q$ case

\vspace*{0.5cm}
\begin{tabular}
[c]{|c|c|}\hline\hline
$\phi_{S}$ & $g_{SV\gamma}$\\\hline
$\phi_{S}\rightarrow a_{0}\gamma$ & $0$\\\hline
$\phi_{S}\rightarrow f_{0}\gamma$ & $\frac{b\cos(\varphi_{S})}{3}$\\\hline
$\phi_{S}\rightarrow\sigma_{B}\gamma$ & $\frac{b\sin(\varphi_{S})}{3}$\\\hline
\end{tabular}
\begin{tabular}
[c]{|c|c|}\hline\hline
$\omega_{N}$ & $g_{SV\gamma}$\\\hline
$a_{0}\rightarrow\omega_{N}\gamma$ & $\frac{b}{2}$\\\hline
$f_{0}\rightarrow\omega_{N}\gamma$ & $-\frac{b\sin(\varphi_{S})}{6}$\\\hline
$\omega_{N}\rightarrow\sigma\gamma$ & $\frac{b\cos(\varphi_{S})}{6}$\\\hline
\end{tabular}
\begin{tabular}
[c]{|c|c|}\hline\hline
$\rho$ & $g_{SV\gamma}$\\\hline
$a_{0}\rightarrow\rho\gamma$ & $\frac{b}{6}$\\\hline
$f_{0}\rightarrow\rho\gamma$ & $-\frac{b\sin(\varphi_{S})}{2}$\\\hline
$\rho\rightarrow\sigma\gamma$ & $\frac{b\cos(\varphi_{S})}{2}$\\\hline
\end{tabular}

\end{center}

\emph{Way 1}: When assuming the mechanism of Fig. 1.a as dominant, a problem
arises: the ratio $\left\vert c_{\phi f_{0}\gamma}/c_{\phi a_{0}\gamma
}\right\vert =\sin(\varphi_{V})/(2\cos(\varphi_{V})\cos(\varphi_{S}%
)-\sin(\varphi_{V})\sin(\varphi_{S}))=2.1$ (from Eq. (\ref{cf})) cannot be
reproduced for a small value of $\varphi_{S}.$ Indeed, a value of $\varphi
_{S}$ close to $\pm90^{\circ}$ is required to explain this ratio: this is in
\emph{clear} disagreement with the result of Eq. (\ref{qfit}) and implies an
unnatural, dominant $\overline{s}s$ content for $f_{0}(600).$ More in details,
the use of Eq. (\ref{cf}) implies $b=-11.8$ GeV$^{-1}$ and $\varphi
_{S}=-99.9^{\circ}.$ As a result, we determine the radiative decays,
summarized in Table 6, which turn out to be too large (note the MeV scale!).
For instance, the value $\Gamma_{a_{0}\rightarrow\omega\gamma}\sim85$ MeV is
clearly unrealistic. Also, $\Gamma_{\rho\rightarrow\sigma\gamma}\sim410$ keV
is incompatible with Ref. \cite{cmd2}.

\bigskip

\begin{center}
\textbf{Table 6: }$S\rightarrow V\gamma$ in the $\overline{q}q$ case using
Ref. \cite{lupod}.%

\begin{tabular}
[c]{|c|c|}\hline\hline
Mode & Decay [MeV]\\\hline\hline
$a_{0}\rightarrow\rho\gamma$ & 10.4\\\hline\hline
$f_{0}\rightarrow\rho\gamma$ & 69.88\\\hline\hline
$\rho\rightarrow\sigma\gamma$ & 0.41\\\hline
\end{tabular}
\begin{tabular}
[c]{|c|c|}\hline\hline
Mode & Decay [MeV]\\\hline\hline
$a_{0}\rightarrow\omega\gamma$ & 85.6\\\hline\hline
$f_{0}\rightarrow\omega\gamma$ & 6.73\\\hline\hline
$\omega\rightarrow\sigma\gamma$ & 0.15\\\hline
\end{tabular}

\end{center}

By using VMD the $\gamma\gamma$ decays read $\Gamma_{f_{0}\rightarrow
\gamma\gamma}\simeq772$ keV$,$ $\Gamma_{a_{0}\rightarrow\gamma\gamma}%
\simeq1767$ keV$,$ $\Gamma_{\sigma\rightarrow\gamma\gamma}\simeq92$ keV, which
are 3 order of magnitudes larger than the values of Eq. (\ref{ggexp}). Varying
$\varphi_{S}$ does not improve the overall situation. There is no need to
discuss this scenario any further: a quarkonium scenario together with a
dominant Fig. 1.a is surely ruled out.

\bigskip

\emph{Way 2}: Using $\Gamma_{a_{0}\rightarrow\gamma\gamma}^{\text{exp}}=0.30$
keV and $\Gamma_{f_{0}\rightarrow\gamma\gamma}^{\text{exp}}=0.29$ keV we
obtain $b_{1}=0.23$ GeV$^{-1}$ and $\varphi_{S}=-20^{\circ}.$ In this case the
mixing angle is in agreement with the strong fit of Eq. (\ref{qfit}). Note,
the corresponding quark-antiquark contribution of the $\sigma$ decay is in
this case $\Gamma_{\sigma\rightarrow\gamma\gamma}=0.13$ keV, thus
small\footnote{A small quark-loop contribution of $\sigma\equiv\sqrt{\frac
{1}{2}}(\overline{u}u+\overline{d}d)$ into $\gamma\gamma$ is also the outcome
of Ref. \cite{sgg}. Namely, the finite dimension -which is a necessary
property of a bound state such as the quarkonium one- is responsible for a
smaller $\gamma\gamma$ result than what a local calculation -which neglects
the finite extension of the $\sigma$ field- would deliver.}. The $S\rightarrow
V\gamma$ results are summarized in Table 7.

\begin{center}
\textbf{Table 7: }$S\rightarrow V\gamma$ in the $\overline{q}q$ case using VMD.%

\begin{tabular}
[c]{|c|c|}\hline\hline
Mode & Decay [keV]\\\hline\hline
$a_{0}\rightarrow\rho\gamma$ & 4.0\\\hline\hline
$f_{0}\rightarrow\rho\gamma$ & 3.3\\\hline\hline
$\rho\rightarrow\sigma\gamma$ & 4.7\\\hline
\end{tabular}
\begin{tabular}
[c]{|c|c|}\hline\hline
Mode & Decay [keV]\\\hline\hline
$a_{0}\rightarrow\omega\gamma$ & 33\\\hline\hline
$f_{0}\rightarrow\omega\gamma$ & 0.61\\\hline\hline
$\omega\rightarrow\sigma\gamma$ & 0.53\\\hline
\end{tabular}

\end{center}

In this case $\rho\rightarrow\sigma\gamma$ and $\omega\rightarrow\sigma\gamma$
are in agreement with the experiment \cite{cmd2}. The order of magnitude is
similar to the one of the values in Table 4. The coupling constants $c_{\phi
f_{0}\gamma}=0.072$ GeV$^{-1}$ and $c_{\phi a_{0}\gamma}=-0.0075$ GeV$^{-1}$
are determined, thus implying that Fig. 1.a is negligible (compare with Eq.
(\ref{cf})). There is however a drawback: $\Gamma_{\phi\rightarrow\sigma
\gamma}=3.28$ keV, which is much larger than the experimental value
$\Gamma_{\phi\rightarrow\sigma\gamma}^{\text{exp}}\lesssim0.6$ keV obtained in
Ref. \cite{aloisio}, and thus seems to be excluded. Such a large $\Gamma
_{\phi\rightarrow\sigma\gamma}=3.28$ keV would probably produce a much more
pronounced $\sigma$ peak, which is however not present in the line shapes
\cite{kloe}.

\section{Discussions and Conclusions}

In this work we studied $S\rightarrow PP$, $S\rightarrow V\gamma$ and
$S\rightarrow\gamma\gamma$ decays and we discussed the connection of the
latter with the radiative process $\phi\rightarrow S\gamma,$ which is subject
of a detailed experimental analysis by the KLOE group.

A general outcome of $S\rightarrow V\gamma$ ratios (Tables 2 and 5), which
does not depend on numerical details, is the predictions of the following
properties. Tetraquark scenario: $\Gamma_{a_{0}\rightarrow\omega\gamma}%
\simeq\Gamma_{f_{0}\rightarrow\rho\gamma}$ and $\Gamma_{a_{0}\rightarrow
\rho\gamma}\simeq\Gamma_{f_{0}\rightarrow\omega\gamma.}$ Quarkonium scenario:
$\Gamma_{a_{0}\rightarrow\omega\gamma}/\Gamma_{a_{0}\rightarrow\rho\gamma
}\simeq\Gamma_{f_{0}\rightarrow\rho\gamma}/\Gamma_{f_{0}\rightarrow
\omega\gamma}\simeq\Gamma_{\rho\rightarrow\sigma\gamma}/\Gamma_{\omega
\rightarrow\sigma\gamma}\simeq9.$

In order to obtain numerical values, two ways have been followed. When
assuming Fig 1.a as the dominant process in the radiative $\phi$ decays
(denoted as way 1, which necessarily implies a preformed tetraquark or
quarkonium substructure of the $f_{0}/a_{0}$ mesons), the results for
$S\rightarrow V\gamma$ are summarized in Tables 3 and 6 in the $\overline
{q}^{2}q^{2}$ and $\overline{q}q$ cases respectively. The $\overline{q}q$ case
is surely excluded because of the unrealistically large values of the
$S\rightarrow V\gamma$ decays (see Table 6). The $\overline{q}^{2}q^{2}$ case
is also disfavored in view of too large rates for $\rho,\omega\rightarrow
\sigma\gamma$ and (VMD deduced) $f_{0}$, $a_{0},$ $\sigma\rightarrow
\gamma\gamma,$ but still not completely ruled out. Experimental result are
needed: if however large decays will be found (such as, for instance,
$\Gamma_{a_{0}\rightarrow\omega\gamma}\gtrsim50$ keV) one could infer that a
compact structure is compulsory. Being the quarkonium interpretation highly
problematic, one would be necessarily left with the tetraquark interpretation
as the only possible one.

When starting from $\gamma\gamma$ data of $f_{0}$ and $a_{0}$ (way 2), the
results, summarized in Tables 4 and 7 for the $\overline{q}^{2}q^{2}$ and
$\overline{q}q$ cases, turn out to be of the same order of the kaon-loop based
calculations \cite{insight,escribano,tanja}. The outcome is in agreement with
the $\rho,\omega\rightarrow\sigma\gamma$ data, but a full and consistent
analysis should then include both direct and meson-loop driven (not considered
here) contributions to $S\rightarrow\gamma\gamma$ and $S\rightarrow V\gamma$
decays (the $\overline{q}q$ case is anyway disfavored because of a too large
$\phi\rightarrow\sigma\gamma$ branching ratio). However, a model-independent
conclusion can be achieved: small $S\rightarrow V\gamma$ radiative decays -if
confirmed experimentally- would surely imply that the $\phi$ decay is
dominated by the kaon loop of Fig. 1.b, independently from the nature of the
scalar states (in agreement with the discussion presented in Ref.
\cite{hanhart}). This, in turn, may allow for a precise determination of the
amplitudes in future updates of the KLOE experiment. It is indeed interesting
to notice that the present strong amplitudes as determined by fits to the line
shapes of Ref. \cite{kloe} assuming the dominance of Fig. 1.b (in GeV), read:
\begin{equation}
\left\vert A_{f_{0}\pi\pi}^{\text{kl-kloe}}\right\vert =1.71\pm0.7,\text{
}\left\vert A_{f_{0}KK}^{\text{kl-kloe}}\right\vert =5.4\pm1.6\text{,
}\left\vert A_{a_{0}\pi\eta}^{\text{kl-kloe}}\right\vert =2.8\pm0.1,\text{
}\left\vert A_{a_{0}KK}^{\text{kl-kloe}}\right\vert =3.06\pm
0.06,\label{amplkloe}%
\end{equation}
and are in rough agreement with the independent results of Eq. (\ref{amplexp}%
), what indeed constitutes a remarkable fact.

We notice that the amplitudes of Eq. (\ref{amplexp}) imply that the tree-level
decays of the $f_{0}$ and $a_{0}$ mesons are large:%
\begin{equation}
\Gamma_{f_{0}\rightarrow\pi\pi}^{\text{tl}}=161\pm25\text{ MeV, }\Gamma
_{a_{0}\rightarrow\pi\eta}^{\text{tl}}=146\pm13\text{ MeV}\label{tl}%
\end{equation}
(which, without inclusion of the $\overline{K}K$ channel, are already larger
than the 50-100 MeV \emph{full} widths reported by PDG\ \cite{pdg}. However,
the PDG widths refer to the peak widths, which are strongly distorted due to
the nearby $\overline{K}K$ threshold. Note, the KLOE result for $f_{0}%
\rightarrow\pi\pi$ is about 60 MeV, but has a large uncertainty, while
$a_{0}\rightarrow\pi\eta$ is $103\pm10$ MeV, which is also sizable.) The very
fact that large decay widths of $f_{0}$ and $a_{0}$ are found is the reason
why a -albeit qualitative- consistent description of a full tetraquark nonet
$\left\{  \sigma,k,f_{0},a_{0}\right\}  $ below 1 GeV is possible, as we
presented in Section 3.2. We recall also that, quite remarkably, such a
consistent description is \emph{not} achieved in the $\overline{q}q$ case: too
small $\sigma$ and $k$ widths are found and the scalar mixing angle is in
disagreement with general arguments based on the $U_{A}(1)$ anomaly. While
small $S\rightarrow V\gamma$ (with $S=f_{0}/a_{0}$) decay widths (as in Table
4) are also in agreement with a loosely bound kaon molecular state, the latter
predict smaller decay widths than Eq. (\ref{tl}): $\Gamma_{a_{0}\rightarrow
\pi\eta}=10$-$60$ MeV and $\Gamma_{f_{0}\rightarrow\pi\pi}\simeq$ $10$-$50$
MeV \cite{tanja,lemmer,lemmernew,isgur}. Indeed, a small width ($\lesssim50$
MeV) is a typical characteristic of a loosely bound kaonic state, as
emphasized in Ref. \cite{isgur}. These predictions are however in disagreement
with Eq. (\ref{tl}). In particular, this criticism holds in the $a_{0}\pi\eta$
channel where the errors are smaller; in the recent work Ref. \cite{lemmernew}
a small width $\Gamma_{a_{0}\rightarrow\pi\eta}$ of at most 30 MeV is found,
which is at odd with both Eq. (\ref{amplexp}) and (\ref{amplkloe}). In view of
this discussion we regard the tetraquark assignment as the most suitable for
the description of light scalar states.

\bigskip

\bigskip
\textbf{Acknowledgments:} We thank Paolo Gauzzi and Cesare Bini for useful
discussions. F. G. thanks BMBF for financial support, G.P. acknowledges
financial support from the Alliance Program of the Helmholtz Association
(HA216/EMMI).

\bigskip

\end{document}